\documentclass[onecolumn,superscriptaddress,aps,pre]{revtex4}
\usepackage{color}
\usepackage{amsmath}
\usepackage{amssymb}
\usepackage{graphicx}
\usepackage{subfigure}
\usepackage{ulem}

\newcommand{\rev}[1]{ {  #1} }

\begin{document}

\title{Growth or Reproduction: Emergence of an Evolutionary Optimal Strategy}

\author{Jacopo Grilli}
\affiliation{Dipartimento di Fisica e Astronomia ``G. Galilei'', Universit\`{a} di Padova, CNISM and INFN, via Marzolo 8, 35131 Padova, Italy}

\author{Samir Suweis}
\affiliation{Dipartimento di Fisica e Astronomia ``G. Galilei'', Universit\`{a} di Padova, CNISM and INFN, via Marzolo 8, 35131 Padova, Italy}

\author{Amos Maritan}
\affiliation{Dipartimento di Fisica e Astronomia ``G. Galilei'', Universit\`{a} di Padova, CNISM and INFN, via Marzolo 8, 35131 Padova, Italy}

\begin{abstract}
\rev{
Modern ecology has re-emphasized the need for a quantitative understanding of the original 'survival of the fittest theme' based on analyzis of the intricate trade-offs between competing evolutionary strategies that characterize the evolution of life. This is key to the understanding of species coexistence and ecosystem diversity under the omnipresent constraint of limited resources.
In this work we propose an agent based model replicating a community of interacting individuals, e.g. plants in a forest, where all are competing for the same finite amount of resources and each competitor is characterized by a specific growth-reproduction strategy. We show that such an evolution dynamics drives the system towards a stationary state characterized by an emergent optimal strategy, which in turn depends on the amount of available resources the ecosystem can rely on. We find that the share of resources used by individuals is power-law distributed with an exponent directly related to the optimal strategy. The model can be further generalized to devise optimal strategies in social and economical interacting systems dynamics.}
\end{abstract}

\pacs{ADD}
\maketitle


\section{Introduction}

\rev{Living systems evolve and adapt to survive in an evolutionary tussle.  One of the main driving forces of natural selection~\cite{Darwin,Wilbur1974} is the competition for resources that regulates survival, growth and reproduction rates. The game of life is orchestrated through an optimization program that releases energy through offspring production while simultaneously trying to conserve it through an evolutionary feedback mechanism.
Growth is the key ingredient to succeed in competition with other individuals, while reproduction promotes colonization through offspring thereby avoiding extinction.

Both empirical and theoretical evidences suggest that a trade-off across species between competition and colonization ability~\cite{Kraaijeveld2001,novak2006,Calcagno2006,Chesson2008} promotes coexistence: the good competitors  are typically bad colonizers whereas the good colonizers are bad competitors.
The essential idea here is borrowed from a well known ecological theory,
the r/K selection theory~\cite{machartur1967,Pianka1970}. This theory states that in ecosystems with access to large (read infinite) resources. the long time dynamics approaches a stable equilibrium with higher growth and lower reproduction rates for the interacting agents (K-strategy). On the contrary, in unstable environments with scarce resources, the individuals procreates more, but have smaller biomasses (r-strategy).
The impact of trade-offs in competition strategies on biodiversity maintenance has also been investigated in more recent theoretical models known as trade-off models~\cite{Muller-landau2002,Kneitel2004}. Both stochastic individual based models \cite{Muller-landau2002,Bertuzzo2011} and resource competition models in the Lotka-Volterra form~\cite{Wangersky1978,Chesson2008} have shown that the inclusion of the trade-offs traits positively affect the maintenance of species diversity within single and multi-trophic ecological communities.
Even in the deterministic limit, excluding all dynamic stochasticity, optimal and evolutionary stable trade-offs have been quantified for a structured rotifer population preying on a dynamically varying food supply~\cite{shertzer2002}.

The present article will focus on an optimal trade-off between growth (through competition) and reproduction (trough colonization) while simultaneously incorporating all inherent stochastic fluctuations due to natural birth, death and growth.}

\section{Definition of the model}

\rev{The model is defined as a birth-death-growth stochastic process in the continuous time limit. It consists of a community of agents competing for a finite amount of resources $R$. Each agent uses a fraction of resources $R$ at any given time, e.g. at the time $t$
the $i$-th individual uses an amount of resources $\epsilon_i(t)$ such that $\sum_i \epsilon_i(t)=R$. In our analysis, we assume $R$ to be a constant parameter, while the number of individuals in the
community $N(t)$ varies during the process and, at stationarity, it fluctuates around its time averaged value.

The population of the community remains unchanged until one of the $N(t)$ individuals dies. Death events occur with a rate $d \ N(t)$, given that each individual has a constant death rate $d$, independent of its history, resource usage or age. The $k$-th individual, which is uniformly drawn from the pool of the $N(t)$ individuals, dies and $\epsilon^\ast:=\epsilon_k(t)$ resources are instantly shared by the rest of individuals in the community.} Individuals use those resources to grow or to have a progeny.  Each individual is characterized by a balance between these two processes, that is quantified in terms a parameter $p_i$.  With probability $p_i$ (or $1-p_i$) the $i$-th individual uses the available energy to produce offspring, i.e. new individuals in the system, or to grow. This complementarity reflects the facts that both growth and reproduction requires energy allocation, and the individuals have to find a balance between the two tasks.

We implement the above growth/birth process in the following way. When an individual dies it frees a quantity of resources $\epsilon^*$. The rest of the community is divided into two disjoint subsets, $B$ and $G$: the $i$-th individual has a probability $p_i$ (or $1-p_i$) to belong to the set $B$ (or $G$), that is,
it chooses the move \textit{birth} (\textit{growth}).
The quantity of freed resources that the $i$-th individual manages to use is proportional to its own energy $\epsilon_i$. This choice is solely decided by the following fact:
bigger an individual is, more resources it will consume in order to survive. Furthermore this hypothesis is consistent with the von Bertalanffy equation  of ontogenetic growth~\cite{Berta57,West2001,Banavar2002}. If we define $E_b=\sum_{i\in B}\epsilon_i\ (E_g=\sum_{i\in G}\epsilon_i)$ as the sum of the energies of the individuals that have chosen the move \textit{birth} (\textit{growth}) then the total energy allocated for the births is $\epsilon^* E_b/(E_g+E_b)$, where the relation $E_g+E_b=R-\epsilon^*$ holds.

Each new generated individual uses a fixed amount of resources $\epsilon^{(0)}>0$. This parameter defines the minimal biomass of an individual.
Given this ``minimal energy'' $\epsilon^{(0)}$, the number $n_b$ of new \rev{individuals} added to the system is given by the largest integer less than $(\epsilon^*/\epsilon^{(0)}) E_b/(E_g+E_b)$, i.e. $n_b=\lfloor(\epsilon^*/\epsilon^{(0)}) E_b/(E_g+E_b)\rfloor$, where $\lfloor\cdot\rfloor$ indicate the floor function. The $p_i$ values of these new individuals are inherited from their parents: for each of the $n_b$ progenies, the $i$-th individual belonging to the subset $B$ has probability $\epsilon_i/E_b$ to transmit its $p_i$.
Individuals belonging to the subset $G$ grow exploiting the remaining energy  $\epsilon_g\equiv\epsilon^*-n_b\epsilon^{(0)}$ (i.e., energy not used for procreation). Therefore, the energy of the $i$-th individual belonging to $G$ changes as $\epsilon_i\rightarrow \epsilon_i(1+\epsilon_g/E_g)$. 

The dynamics defined above conserves the total resource $R$ of the ecosystem, but do not conserve the total number of individuals in the community. If the fluctuations of the total population $N$ drive the system to the state $N=0$, then the dynamics stops. If the system reaches the $N=0$ state, we initialize the system to a new configuration with $R/\epsilon^{(0)}$ individuals (of energy $\epsilon^{(0)}$). This choice may be interpreted as immigrations of outside species in the the community. The proposed community dynamics also has ``non-trivial'' absorbing states. These are related to the strategy $p$: if all the individuals share the same strategy $p$ there is no way to introduce a new individual with a different $p$ in the ecosystem. Therefore the stationary state of the ecosystem dynamics will be characterized by one final strategy $p$ and not by a mixture of heterogeneous $p_i$. From a theoretical point of view the number of these possible absorbing states is infinite (one for each value of $p$). 

\section{Fixed $p$}

\subsection{Numerical simulations}

The simulations of the community dynamics are performed using Gillespie's algorithm. The system is initialized at time $t=0$ with $N(0)=R$ individuals having the same energy usage ($\epsilon^{(0)}$=1) and the same value of $p$. The structure of the community depends on $R$ and $p$. The main goal of the simulations is to study this dependence.

Suppose that at time $t$ there are $N(t)$ individuals in the ecosystems and that a death event occurs at time $\tau>t$.
Each individual has a constant death rate $d$, and therefore $\tau$ is a random variable with exponential distribution and average $1/dN(t)$. At time $\tau$, we randomly remove an agent that frees its resource $\epsilon^\ast$. The remaining individuals identically compete to gain their shares of the released resources.

\begin{figure*}[tbp]
\centering
  \includegraphics{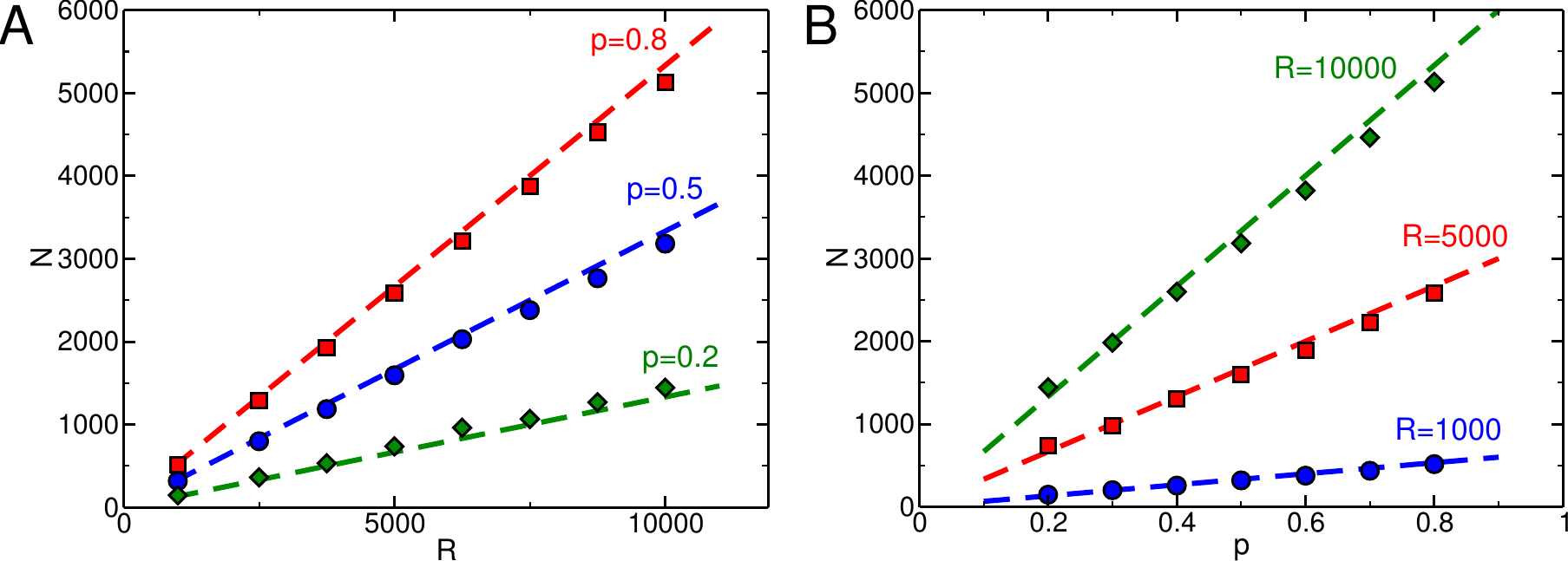}
  \caption{
	Relation between average number of individuals $\big<N\big>$, total available resources
	$R$ and growth/reproduction strategy $p$ under the proposed ecosystem dynamics modeling. Numerical simulations (points of different colors/shapes)
	show a linear relation of the form $\big<N\big>=b p R/\epsilon^{(0)}$, with a $b =0.64 \pm 0.02$.
	The simulation were performed with $\epsilon^{(0)}=1$ and averages were performed over $100$ realizations. 	
	Lines represent analytical mean-field prediction $\big<N\big>=(2/3)p R/\epsilon^{(0)}$
	obtained in section~\ref{sec:mf}, where the proportionality constant is predicted to be $b=2/3$.
	}
\label{fig:pop}
\end{figure*}

The number of individuals, $N(t)$, is a random variable, whose average at stationarity depends on both $R$ and $p$. Numerical simulations, shown in Fig.~\ref{fig:pop}, indicates that $\big< N \big> = b p R$ with $b=0.64\pm0.02$, which is compatible with our analytical estimate $b=2/3$ given in section~\ref{sec:mf}.

\begin{figure*}[tbp]
\centering
  \includegraphics{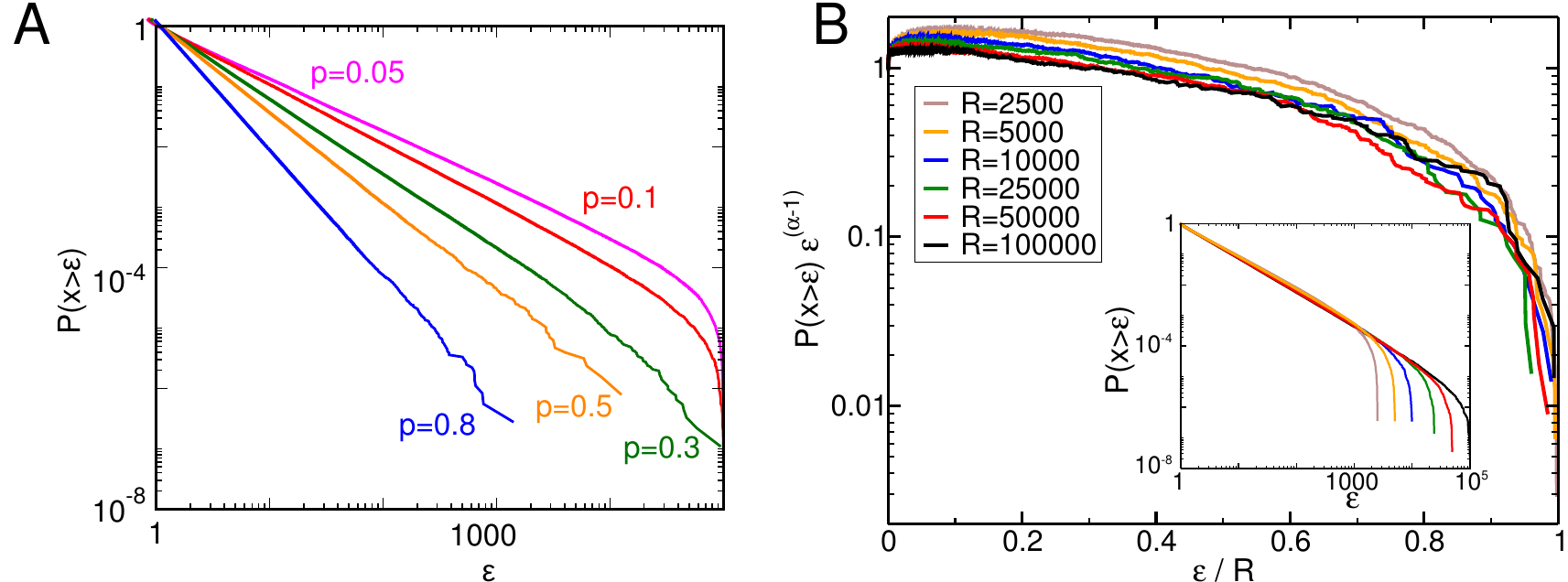}
  \caption{Power-law behavior of $n(\epsilon|p,R)$.
	Panel A 
	shows the cumulative distribution of resource usage. i.e.
	the average fraction of individual using at least an amount $\epsilon$ of resources,
	$P(x>\epsilon)=\int_{\epsilon}^{\infty} n(x|p,R) \mathrm{d} x/\int_{0}^{\infty}
	n(x|p,R) \mathrm{d} x$, for $R=10^5$ and different values of $p$. It turns out to be
	a truncated power-law, with an exponent $-\alpha(p)+1$.
	Panel B shows the
	collapse of distributions obtained for different values of $R$ at fixed $p=0.2$. We plot
	$\epsilon^{\alpha(p)-1} P(x>\epsilon)$ vs. $\epsilon/R$ , where
	$\alpha(p)=2.16$  and for different values of $R$. The collapse is consistent with our
	finite size scaling hypothesis, i.e. $n(\epsilon|p,R)=\epsilon^{-\alpha(p)}f(\epsilon/R)$.
	The inset shows $P(x>\epsilon)$ for different values of $R$.
	Notice the scale in the vertical axes and in the inset. 
	In both panels $\epsilon^{(0)}=1$. Averages were performed over $100$ realizations.
	}
\label{fig:powerlaw}
\end{figure*}

The community is characterized in terms of the distribution of resource usage among individuals as measured by  $n(\epsilon|p,R)\mathrm{d}\epsilon$, the average number of individuals in the energy interval $(\epsilon,\ \epsilon+\mathrm{d}\epsilon)$ when all the individuals share the same value of $p$. Figure~\ref{fig:powerlaw}A shows that resources are distributed with a truncated power-law with an exponent $-\alpha(p)$ depending on $p$.
As expected smaller the values of $p$ is (competition strategy biased favorably), smaller is the absolute value of exponent of the distribution.

The cut-off of the distribution of order $R$ suggests that finite size scaling~\cite{Fisher1972} is a good candidate to describe the behavior of the resource usage in the community. This corresponds to the assumption that $n(\epsilon|p,R)=\epsilon^{\alpha(p)}f(\epsilon/R)$, where $f(x)$ tends to a constant value for small arguments whereas it rapidly approaches zero when the argument increases. If finite size scaling holds the plot of $\epsilon^{\alpha(p)}n(\epsilon|p,R)$ (or equivalently that of $\epsilon^{\alpha(p)-1}\int_\epsilon^\infty n(x|p,R)\mathrm{d}x$ for a better statistics) versus $\epsilon/R$ should give a single data collapsed curve instead of multiple curves for different values of $R$. The quality of the collapse we get (see Fig.~\ref{fig:powerlaw}B) is consistent with the finite size scaling ansatz.

\subsection{Mean-field approximation}
\label{sec:mf}

\begin{figure*}[tbp]
\centering
  \includegraphics{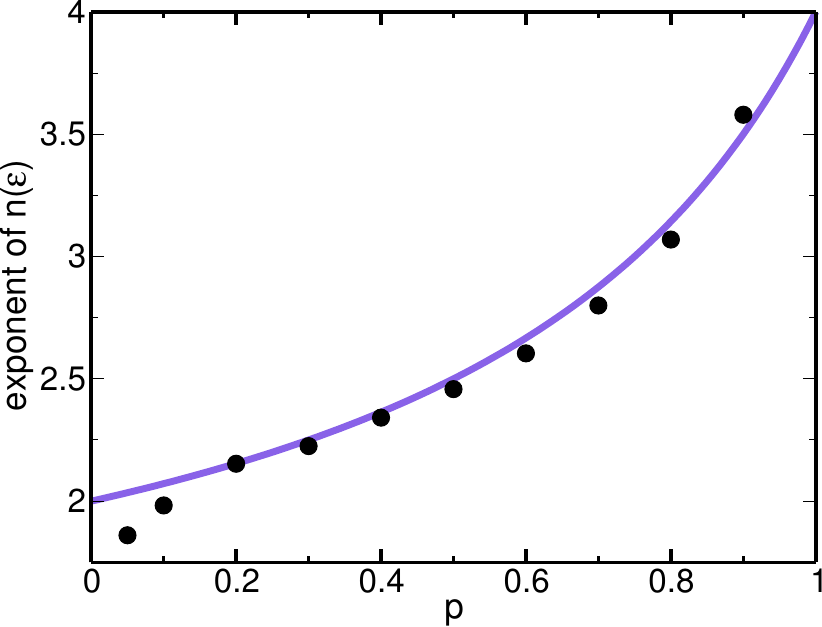}
  \caption{
	Comparison between the exponent $\alpha(p)$ obtained by fitting simulation data with a power-law (black dots) and our analytical prediction $\alpha=(6-2p)/(3-2p)$ (purple curve)
	based on the mean-field approximation.
	}
\label{fig:exponent}
\end{figure*}

In this section, we perform a mean field approximation in order to derive both the power law decay and an estimate of the decay exponent as a function of $p$. The time evolution of the energy usage distribution in the mean field approximation is given by (from now on we shall omit the $p$ and $R$ dependence in $n(\epsilon|p,R)$ in order to simplify the notation)
\begin{equation}
\begin{split}
\frac{\partial n(\epsilon,t)}{\partial t} & =
- d \ n(\epsilon,t)+ d \ \big<N(t)\big> \big<n_b\big> \delta(\epsilon-\epsilon^{(0)}) +  \\
& d \ \big<N(t)\big> (1-p) \Bigl[  \frac{1}{D} n\big(\frac{\epsilon}{D},t\big) - n(\epsilon,t)
 \Bigr]
 \ ,
\end{split}
\label{eq:meanfield}
\end{equation}
where the first term represents the decay of $n(\epsilon,t)$ due to random death events, the second term takes into account the birth events, which are $\big<n_b\big>$ on average, while the third term represents growth.
$\big<N(t)\big>$ is the total number of individuals (i.e. $\big<N(t)\big>=\int d \epsilon n(\epsilon,t)$) and $D\epsilon$ is the updated energy usage of all individuals belonging to the growing set which originally had a resources usage equal to $\epsilon$ with
\begin{equation}
D = 1 + \frac{\big<\epsilon^*\big>-\big< n_b \big> \epsilon^{(0)}}{(1-p)(R-\big<\epsilon^*\big>)} \ .
\label{eq:Dgrowth}
\end{equation}
From equations~\ref{eq:meanfield} and~\ref{eq:Dgrowth} one has $D(\epsilon^*)=\big< 1+\epsilon_g/E_g\big>= 1+\big<(\epsilon^*-n_b\epsilon^{(0)})/E_g\big>\approx 1+ (\big<\epsilon^*\big>-\big<n_b\big>\epsilon^{(0)})/\big<E_g\big>$ where $\big<E_g\big>=(1-p)(R-\big<\epsilon^*\big>)$.
The amount of freed resources at time $t$ depends on the probability that an individual with a resource usage $\epsilon^*$ dies. Since the individual chosen to die is randomly selected, we have $\big<\epsilon^*\big>= \big< \int \mathrm{d} \epsilon n(\epsilon,t)\epsilon/N \big> \approx R/\big<N\big>$.
At stationarity $n(\epsilon,t)=n(\epsilon)$ and, therefore,
\begin{equation}
\epsilon \frac{\partial n(\epsilon)}{\partial \epsilon} = -
\frac{2-b}{1-b}n(\epsilon) + \frac{b}{1-b} \frac{R}{\epsilon^{(0)}} \delta(\epsilon-\epsilon^{(0)})
 \ ,
\label{eq:stationary}
\end{equation}
where
\begin{equation}
b = \big< n_b \big> \epsilon^{(0)} \big<N\big>/R
 \ ,
\label{eq:stationary}
\end{equation}
here we assumed that $ \big< \epsilon^*\big>/R \sim 0  $ ,i.e. $|D-1|\sim 0$, which allows us to expand $n(\epsilon/D,t)$ in eq.~(\ref{eq:meanfield})
in order to get eq.~\ref{eq:stationary}.
The solution of this equation for $\epsilon > \epsilon^{(0)}$ is the power law
\begin{equation}
 n(\epsilon) = \frac{1}{\epsilon^{(0)}} \frac{b}{1-b}
\frac{R}{\epsilon^{(0)}} \Bigl( \frac{\epsilon}{\epsilon^{(0)}} \Bigr)^{-\alpha} \Theta(\epsilon-\epsilon^{(0)}), \quad \alpha=\frac{2-b}{1-b}
 \ .
\label{eq:stationary}
\end{equation}
This solution is consistent since $\int d  \epsilon \ n(\epsilon) \ \epsilon=R$ without having to explicitly impose it, whereas  $\big<N\big>=\int d \epsilon \ n(\epsilon)=Rb/\epsilon^{(0)}$ implying that $b = \epsilon^{(0)} \big<N\big>/R$ and so $\big< n_b \big>=1$. Since $\big< n_b \big>$ is the average number of births per dead individual, at stationarity $\big< n_b \big> = 1$ assures a community of constant average size. 
On the other hand the average value of $n_b$ can be explicitly written as
\begin{eqnarray}
  \big<n_b\big> &=& \int d\epsilon^* \Bigl<\frac{n(\epsilon^*)}{N}\Bigr> \Bigl\lfloor \frac{p\epsilon^*}{\epsilon^{(0)}}\frac{R}{R-\epsilon^*} \Bigr\rfloor \approx \int d \epsilon^* \frac{n(\epsilon^*)}{\big<N\big>} \Bigl\lfloor \frac{p\epsilon^*}{\epsilon^{(0)}} \Bigr\rfloor \nonumber \\
   &=&  \frac{1}{\big<N\big>} \frac{p}{\epsilon^{(0)}} \sum_{i=1}^\infty i \frac{\epsilon^{(0)}}{p} \int_{\frac{i}{p}\epsilon^{(0)}}^{\frac{i+1}{p}\epsilon^{(0)}}
d \epsilon^* n(\epsilon^*) \ .
\label{eq:nb}
\end{eqnarray}
If we simply remove the floor function from the above expression we would obtain $b(p)=p$. In order to get a better approximation we estimate the series in eq.~(\ref{eq:nb}) by evaluating the total resource, $R$, as follows
\begin{equation}\label{eq:mid-point}
R=\int d\epsilon \ n(\epsilon) \ \epsilon \approx \sum_{i=1}^\infty (i+\frac{1}{2}) \frac{\epsilon^{(0)}}{p} \int_{\frac{i}{p}\epsilon^{(0)}}^{\frac{i+1}{p}\epsilon^{(0)}} d \epsilon \
n(\epsilon)
\end{equation}
where we have use the ``mid-point prescription'' to approximate the first term in the integrand, $\epsilon$, in the interval $(i\epsilon^{(0)}/p,\ (i+1)\epsilon^{(0)}/p)$.
Using the previous two equations and the definition of $\big<N\big>$, we get $R=\big<N\big>\epsilon^{(0)}(\big<n_b\big>+1/2)/p= 3\big<N\big>\epsilon^{(0)}/(2p)$ which leads to $b=2p/3$. Therefore we predict the exponent of the power-law distribution to be equal to $(6-2p)/(3-2p)$. Figure~\ref{fig:exponent} shows the analytical prediction of the exponent versus the results of numerical simulation. Apart from a mismatch for very small values of $p$ the agreement is quite satisfactory.
This mismatch is due to the fact that the
smaller the value of $p$ is, the bigger is $\big<\epsilon^*\big>\approx R/\big<N\big> = 3\epsilon^{(0)}/2 p$, which we assumed to be very small with respect to $R$. Moreover when $p$ is small, the integration interval in eq.~(\ref{eq:nb}) is wide, and the ``mid-point prescription'' becomes less accurate.
\rev{The mid-point prescription used in eq.~(\ref{eq:nb}) fails if $p>1 - 1/N$ and a better approximation is obtained by removing the floor function.}

\section{Evolutionary strategy: optimal $p$ as a function of the resources}

\begin{figure}[tbp]
\centering
  \includegraphics{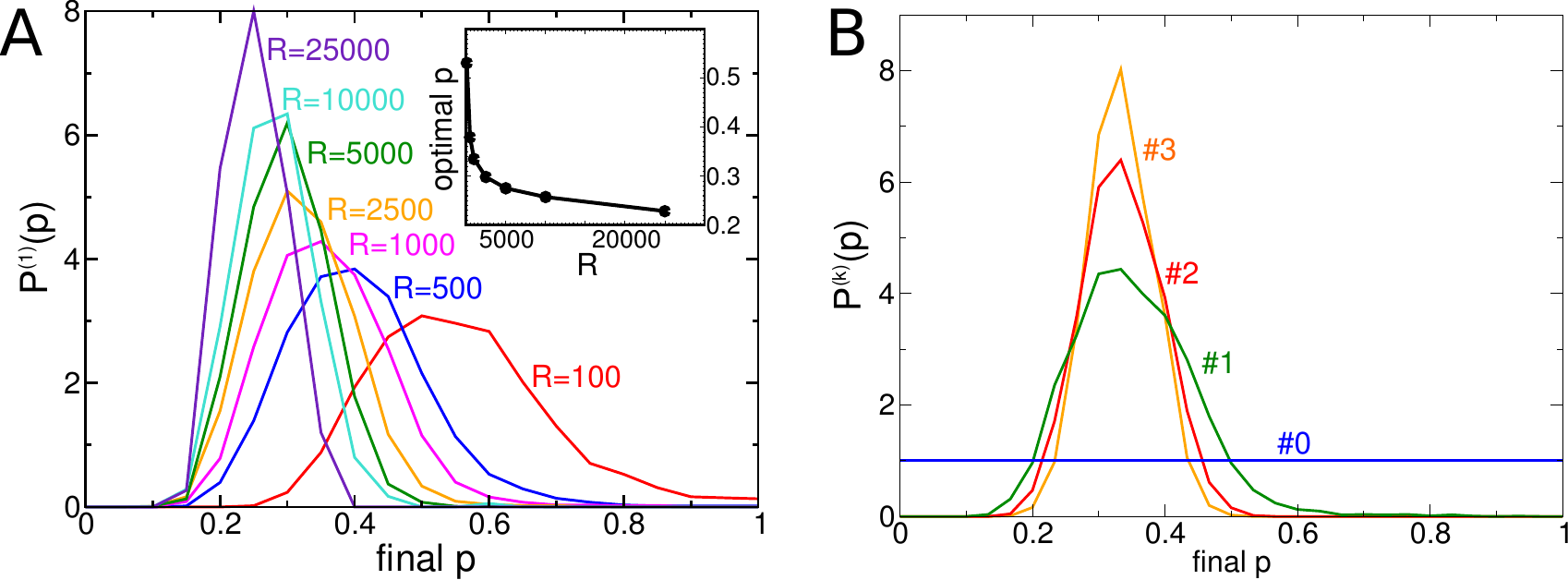}
  \caption{Emergence of the optimal strategy.
	Panel A shows the final distribution of $p$-values, $P^{(1)}(p)$, starting from a uniform distribution
	 $P^{(0)}(p)$ of initial strategies. The distribution
	is peaked around the optimal value which in turn depends on the amount of resources, $R$.
	The inset shows the optimal value of $p$, $p^*(R)$, as a function of the total amount
	of resources, $R$. $p^*(R)$ is evaluated from the peak of distribution $P^{(1)}(p)$.
	Panel B shows the iteration of the optimization procedure in the case of $R=1000$: starting from a uniform distribution $P^{(0)}(p)$
	we get $P^{(1)}(p)$. We then re-initialize the system with a distribution $P^{(1)}(p)$ and obtain $P^{(2)}(p)$ and so on.
	This distribution becomes narrower and narrower, which eventually converges to a delta function centered about $p^*(R)$, showing the existence of
	an optimal strategy.}
\label{fig:finalp}
\end{figure}

The next step is to investigate the fate of an inhomogeneous system with randomly drawn $p$'s for the individuals in the community. As already observed, the system reaches a state with a unique value of $p$. However given that the dynamics is stochastic the selected final $p$ is not unique even in the case of the system starting from the same initial state. Figure~\ref{fig:finalp}A shows the distribution of the final values of $p$, $P^{(1)}(p)$ when we initialize the system with $N_0$ individuals each with an energy usage $\epsilon^{(0)}$ ( $R=N_0\epsilon^{(0)}$), and a uniform distribution of $p$'s, $P^{(0)}(p)$, between $0$ and $1$. Quite interestingly $P^{(1)}(p)$ is not uniform as the initial distribution, $P^{(0)}(p)$, but rather it is peaked around a certain value, $p^*(R)$,  that depends on $R$. Indeed, the center of the distribution gives the optimal $p^*(R)$ value. In fact, by drawing the initial values of $p$ from $P^{(1)}(p)$, we obtain a narrower distribution, $P^{(2)}(p)$ (see Fig.~\ref{fig:finalp}B), and for each iteration the distribution becomes narrower and narrower and centered around $p^*(R)$. In the limit of  $S\rightarrow\infty$ steps, $P^{(S)}(p)$ converges to a delta function. This result is independent of the initial distribution we choose as far as it is different from zero at $p^*(R)$.
Therefore the optimal strategy $p^*(R)$  is evolutionary selected by the community dynamics depending on the ecosystem total resources $R$.
If $R$ is small the population sizes are also small and demographic fluctuations can lead to rapid extinction. In this condition large values of $p$ are selected, favoring procreation with respect to individual growth, and thus promoting the persistence of the ecosystem population. On the other hand, if the available resources are large, demographic fluctuations are less dangerous and species develop a strategy in order to gain the highest share of the available resources, i.e. they prefer to grow and small $p$ strategies are favored.

\begin{figure*}[tbp]
\centering
  \includegraphics[width=0.9\textwidth]{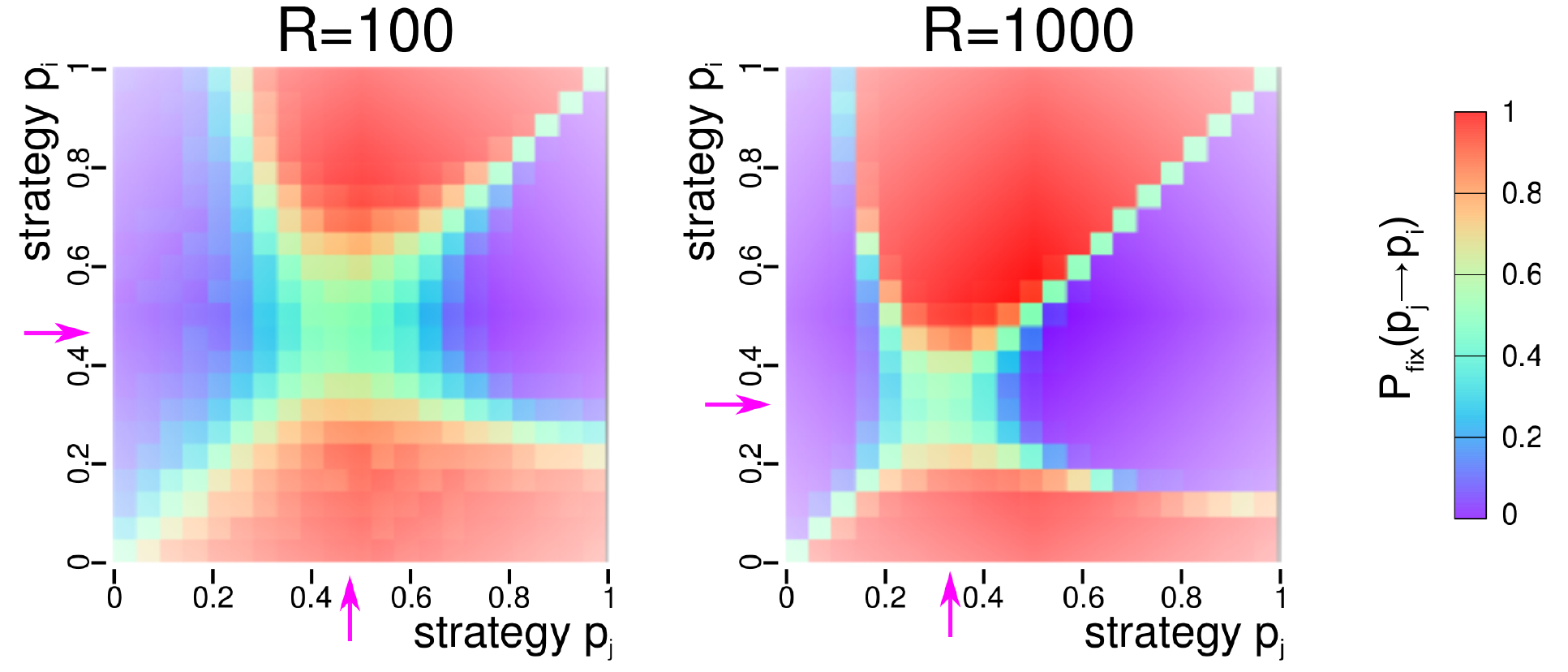}
  \caption{ 
  Probability of extinction and fixation of two competive strategies. In the initial state there are only two species equally populated, one with a strategy $p_i$ and the other one with $p_j$. The element in the $i$-th row and $j$-th column of the plotted matrices represents the probability $P_{fix}(p_j\to p_i)$ that the species with the $p_i$ becomes extinct whereas the one with $p_j$ spreads through all the system ($i,\ j=1,\dots, 20$). The optimal stratgy (ESS) corresponds to the choice  of $p$ which has a probability greater than $1/2$ to be fixed against any other possible value of $p$. The value of $p^*(R)$ depends on the quantity of available resources $R$ and it is plotted in the inset of Fig.~\ref{fig:finalp}A.}
\label{fig:fight}
\end{figure*}

\rev{A natural follow-up of the previous analysis is an attempt to understand whether $p^*(R)$ is an evolutionary stable strategy (ESS). If a population adopts the ESS, it cannot be invaded by a population with another strategy.
In the context of a stochastic dynamics every strategy can in principle be invaded, even if it is the optimal one, just because of demographic fluctuations. The ESS is thus defined as the strategy whose probability to invade another strategy is always higher than the probability to be invaded.  In other words, given two strategies, say $p^*$ and $p$, $p^*$ is the ESS if
\begin{equation}\label{eq:p_fix2}
\displaystyle
P_{fix}(p^*\to p|n^{(0)},m^{(0)}) > P_{fix}(p\to p^\ast|n^{(0)},m^{(0)}) \ \ \forall p \neq p^\ast \ ,
\end{equation}
where, given any two strategies $p_1$ and $p_2$, $P_{fix}(p_1 \to p_2|n,m)=\lim_{t \to \infty} P(n>0,m=0,t|n=n^{(0)},m=m^{(0)},t=0)$ and  $P(n>0,m=0,t|n=n^{(0)},m=m^{(0)},t=0)$ is the probability that starting at $t=0$ with $n^{(0)}$ individuals with $p_1$ and $m^{(0)}$ individuals with $p_2$, at time $t$ all individuals adopt the $p_1$ strategy. The existence of an ESS and its value is independent of the particular choice for $n^{(0)}$ and $m^{(0)}$. We set $n^{(0)}=m^{(0)}$. Figure~\ref{fig:fight} highlights that $p^*(R)$ is indeed an evolutionary stable strategy.}

\section{Discussion and Conclusions}

\rev{Our stochastic birth-death-growth model rationalizes the emergence of an optimal trade-off between growth (competition) and reproduction (colonization) strategies in evolving living systems relying on a finite pool of resources. The dynamical evolution of the system spontaneously selects the optimal competition/colonization strategy to best cope with the environment: if the pool of available resources is small, procreation is deemed a better strategy with respect to individual growth; on the other hand, if the available resources are large, growth (competition) is favored.  Our results are in agreement with empirical evidences: for instance, it has been observed~\cite{mueller1991,freitas2005density} that in populations with high densities of individuals competition abilities are favored, while when the density of individuals is low, much more effort is dedicated towards reproduction.

Our model also predicts a power-law distribution of resources among individuals, and allows us to relate it with the optimal trade-off strategy for the system. Indeed, the exponent of the power-law characterizing the resources distribution depends on the trade-off between growth and reproduction and therefore it is directly related to the availability of resources.
Power-law emerges because of the ``rich get richer'' feature of the individuals growth, and therefore it is not unexpected since preferential attachment mechanisms are known to produce power-laws. Indeed, scale-free distributions of resources among individuals are found in several different contexts: in social systems the wealth among individuals is usually described by a Pareto distribution~\cite{Kleiber2003}. In the context of economy the sizes and incomes of companies follow the Zipf's power law~\cite{Okuyama1999,Axtell2001}; In forest science power-law 
distribution of plants height - which correlates with resources usage - has also been recently reported~\cite{Simini2010}. Interestingly, though, usually these resource distributions show non-universal power law exponents\cite{Baek2011}. For instance distributions of firm sizes have exponents that depend on the firm' sector type~\cite{Axtell2001} or the distribution of family names has a power law decay with an exponent varying from $1.83$ in the US  to about $1$ in Korea \cite{Baek2011}. 
Therefore  our theoretical framework can be properly generalized to model systems also in these different fields of science and to elucidate connection between the exponent of the power-law distributions of resources and the optimal trade-off strategy, which in turn depends on the quantity of available resources.

The proposed model is just the starting point for more complicated dynamics where more complex ingredients may be incorporated. In ecology space plays an important role, as  all individuals compete within a certain area. Therefore the proposed model can be seen as a mean-field version of a spatially explicit model, where the quantity of resources $R$ plays the role of an effective parameter that quantifies the amount of resources and space one individual typically competes for. Similarly space plays a key role also in sociological and economical systems, where the network determining agents specific interactions may display complex topological properties.}

\section{Acknowledgements}
We thank T.~Anfodillo, J.R.~Banavar, C.~Borile, F.~Simini and L.~Tamburino for
insightful discussions and Cariparo foundation for financial support. We are particularly thankful
to S.M.~Bhattacharjee and A.K.~Chattopadhyay for a very helpful critical reading of the manuscript.

\bibliographystyle{unsrt}
\bibliography{biblio}

\newpage

\end{document}